\documentclass[conference]{IEEEtran}
\IEEEoverridecommandlockouts
\pdfoutput=1
\usepackage{cite}
\usepackage{amsmath,amssymb,amsfonts}
\usepackage{algorithm}
\usepackage{algorithmic}

\usepackage{graphicx}
\ifCLASSOPTIONcompsoc
    \usepackage[caption=false, font=normalsize, labelfont=sf, textfont=sf]{subfig}
\else
\usepackage[caption=false, font=footnotesize]{subfig}
\fi
\usepackage{tabularx}
\usepackage{multirow}
\usepackage{url}
\usepackage{textcomp}
\usepackage{xcolor}
\def\BibTeX{{\rm B\kern-.05em{\sc i\kern-.025em b}\kern-.08em
    T\kern-.1667em\lower.7ex\hbox{E}\kern-.125emX}}
\usepackage{multirow}

\def\BibTeX{{\rm B\kern-.05em{\sc i\kern-.025em b}\kern-.08em
    T\kern-.1667em\lower.7ex\hbox{E}\kern-.125emX}}
    
\begin{document}
\bstctlcite{IEEEexample:BSTcontrol}

\title{COCOA: Cold Start Aware Capacity Planning for Function-as-a-Service Platforms}
\author{
\IEEEauthorblockN{Alim Ul Gias}
\IEEEauthorblockA{\textit{Department of Computing} \\
\textit{Imperial College London}\\
London, UK \\
a.gias17@imperial.ac.uk}
\and
\IEEEauthorblockN{Giuliano Casale\thanks{A. Gias is a commonwealth scholar, funded by the UK government. The work of G. Casale is partially supported by RADON, funded by the European Union's Horizon 2020 research and innovation program under grant agreement No. 825040.}}
\IEEEauthorblockA{\textit{Department of Computing} \\
\textit{Imperial College London}\\
London, UK \\
g.casale@imperial.ac.uk}
}

\maketitle
\begin{abstract}
Function-as-a-Service (FaaS) is increasingly popular in the software industry due to the implied cost-savings in event-driven workloads and its synergy with DevOps. To size an on-premise FaaS platform, it is important to estimate the required CPU and memory capacity to serve the expected loads. Given the service-level agreements, it is however challenging to take the cold start issue into account during the sizing process. We have investigated the similarity of this problem with the hit rate improvement problem in TTL caches and concluded that solutions for TTL cache, although potentially applicable, lead to over-provisioning in FaaS. Thus, we propose a novel approach, COCOA, to solve this issue. COCOA uses a queueing-based approach to assess the effect of cold starts on FaaS response times. It also considers different memory consumption values depending on whether the function is idle or in execution. Using an event-driven FaaS simulator, \textit{FaasSim}, we have developed, we show that COCOA can reduce over-provisioning by over 70\% in some workloads, while satisfying the service-level agreements.
\end{abstract}

\begin{IEEEkeywords}
Function-as-a-service, serverless computing, cold start, sizing, layered queueing network
\end{IEEEkeywords}

\section{Introduction}
\label{sec:intro}

Function-as-a-Service (FaaS) platforms, based on the serverless execution model \cite{wang2018peeking}, allow developers to deploy their codes as individual functions without having to deal with the underlying infrastructure management. This facilitates DevOps practices \cite{jabbari2018towards} by providing more flexibility to each development team and increasing the pace of delivery of code updates. The availability of open source platforms, like OpenFaaS and OpenLambda, has made it possible to install on-premise FaaS platforms, which calls for dedicated sizing and resource allocation methods in order to meeting service-level agreements (SLAs). 

FaaS platforms are designed to implement event-driven applications, which react to a change of state as a result of events generated by the environment and execute associated business logic. In a FaaS platform, this logic is termed as a function, which is usually packaged as a container. To reduce resource wastage, FaaS containers are offloaded from the memory given that they remain idle for specific time period. When a new request for an offloaded function arrives, the request is blocked until the function is loaded again. This issue is known as the {\em cold start} issue~\cite{wang2018peeking,lloyd2018serverless}.

During the capacity planning process, the cold start issue can pose a significant trade-off between latency and memory allocation optimization. A cold start occurs when a function is invoked while the corresponding container is not yet loaded in memory, which adds a delay in sending the response needed to spin-up the container and the function runtime dependencies. Despite hurting performance, this mechanism aims at reducing memory consumption by offloading functions that are idle for a sufficiently long time. This poses a trade-off between response time SLAs and available memory to support concurrent execution of more functions, which needs to be considered upon sizing an on-premise installation.

To address this issue, we can draw parallels between a FaaS platform and a Time to Live (TTL) cache. Similar to a FaaS platform, a TTL caching system also periodically offloads its cached objects, so that the cold start issue resembles the object hit rate improvement problem in a TTL cache, in which one needs to decide on the optimal time to keep objects in cache \cite{basu2018adaptive}. Thus, analysis methods  from TTL cache research such as the characteristic time approximation in~\cite{che2002hierarchical} may be in principle applicable also to FaaS sizing in order to estimate the required memory capacity. However, from our study we have identified two limitations of such an approach. First, contrary to TTL cache misses, the latency incurred by function cold start times can vary widely. Next, while a large fraction of TTL research considers equal-sized objects, a function consumes different amount of memory depending on whether it is idle or in execution. 

In this paper, we present COCOA, a sizing method that leverages a stochastic modeling approach based on layered queueing networks (LQN) \cite{franks2008enhanced} and M/G/1-type queueing systems for capacity prediction. To consider the effect of cold starts, we have incorporated the probability of experiencing cold starts, by each function, with the LQN model. These probabilities are estimated from an M/M/1/setup/delayedoff model, which is a variant of M/M/k setup class of models \cite{gandhi2013exact}, which we solve using matrix-analytic methods as a special case of M/G/1-type system. Setup models can approximate the cold start probability for a function, taking into account cold-starts. To predict the required capacity, COCOA follows an iterative process. It repeatedly solves the LQN model to find a set of function idle times and a CPU configuration such that the function response times are just below the SLA. To accelerate the searching process, we have designed a parallel algorithm, where each parallel branch utilizes binary search. 
Once the idle times and CPU configuration are obtained, COCOA estimates the CPU utilization value for each function. These estimations are integrated with a capacity estimation method for TTL cache \cite{fricker2012versatile} to predict the required memory capacity.

Overall, we summarize our contributions as follows: 
\begin{itemize}
    \item We investigate in Section II the similarity between a FaaS platform and TTL cache from the cold start perspective and illustrate that TTL cache analysis, despite promising, is alone insufficient for FaaS capacity estimation.
    \item We present in Section III an LQN-based performance modeling technique for FaaS platform that captures the effect of cold starts over function response times, correcting the limitations of TTL cache analysis when applied to this setting.
    \item In Section IV we propose COCOA, a sizing method for on-premise FaaS platforms leveraging our LQN model and demonstrate its effectiveness in reducing resource over-provisioning while ensuring response time SLAs.
\end{itemize}
lastly, in Section V we validate our framework against data from simulation. Sections VI and VII respectively position the work against the state-of-the-art and conclude the paper.


\section{Similarity between FaaS and TTL caches}
\label{sec:analogy}
\subsection{Analogy}
\label{subsec:analogy}
TTL caches, used in content delivery networks, allow faster page loading and reduction of load at the origin server. In such caching systems, each cache object is associated with a TTL, after which the object is evicted \cite{basu2018adaptive}. If the cache can serve the request for an object, it is termed as a cache hit. The fraction of request served, for a particular period, is called the hit rate. Longer TTL values can improve the hit rates but are more costly as they require a larger cache size.

Similarly to TTL caches, to reduce the number of cold starts, a possible solution, from the point of view of the end user, is to keep the functions in the memory for longer periods. 
However, this significantly increases the required memory capacity since most of the functions always remain loaded. A way around this problem could be to determine an optimal idle time that ensures a certain degree of availability and reduce the number of cold starts to an acceptable limit. We notice the analogy of this problem with the configuration of TTL caches \cite{basu2018adaptive}. In such systems, the goal is to determine a characteristic time \cite{che2002hierarchical}, which is set as the TTL value, that maximizes the cache hit rates with a given space constraint. The hit rate ($h_i$) for an object is defined by (\ref{eq:che-hit-rate}), considering its TTL ($T_i$) value is reset upon a new request \cite{dehghan2019utility}. 

\begin{IEEEeqnarray}{CC}
h_i= 1-e^{-\lambda_iT_i}
\label{eq:che-hit-rate}
\end{IEEEeqnarray}

This TTL value is similar to the idle time of the functions. It represents a period for which a object is kept in the cache, even though no new request is received,  which ensures a certain degree of availability. Similarly for FaaS, an idle time represent a period where the functions are kept loaded in the memory even though they are idle. Therefore, we can use this concept of TTL to estimate the idle times for the functions. To realize this, we can simply solve (\ref{eq:che-hit-rate}) to get a TTL value for a particular hit rate. This value can be set as the idle time of the function. It will ensure the needed degree of availability and reduce the number of cold starts. Consequently, this will help to satisfy the response time constraints. 



\subsection{Example}

To illustrate the concept, we have developed a discrete-event simulator for a FaaS platform, referred to in the rest of the paper as \textit{FaasSim}\footnote{The simulator is available for download at -  \url{https://github.com/alimulgias/FaasSim}}. Developing \textit{FaasSim} was necessary since popular performance modeling tools, like JMT \cite{bertoli2009jmt}, cannot model the cases we need to consider for FaaS - the cold starts and modeling both CPU and memory consumption.

In the simulation, we have considered an open workload model where the requests arrive following a Poisson process. To introduce popularity among the functions, meaning their invocation probabilities will be different, we have used the Zipf distribution \cite{glassman1994caching}. The function service times are set such that they are at-most half of the SLA value of 2 seconds, when there is no resource contention. The cold start times are chosen from a recent study on popular FaaS platforms \cite{mikhail2019comparison} that, apart from the platform, also considered factors like programming languages and deployment sizes, which effect the magnitude of cold starts.

\begin{figure}[t] 
    \centering
  \subfloat[Hit rates and response time \label{fig:motivation-response}]{%
      \includegraphics[width=0.5\linewidth]{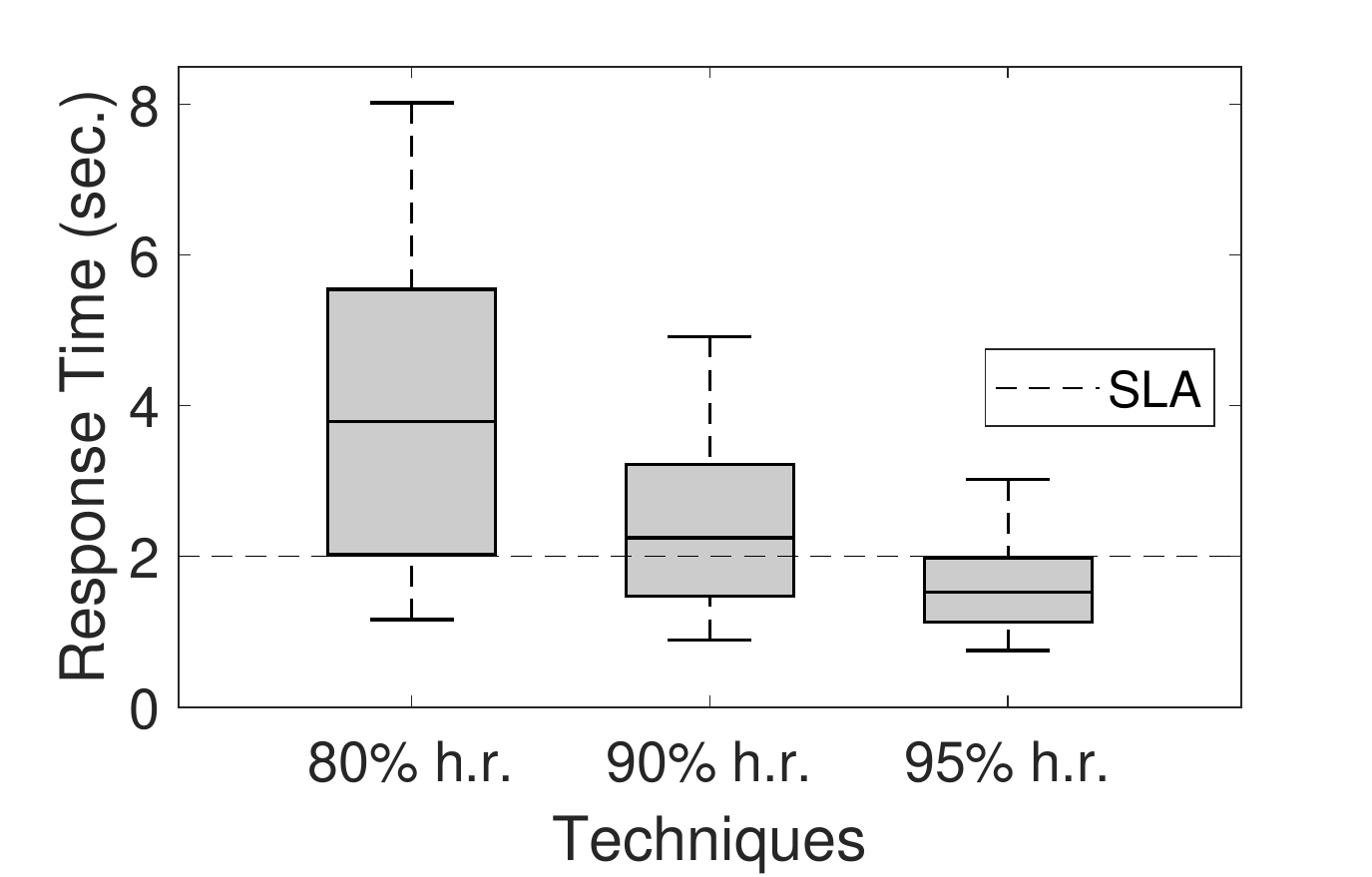}}
  \subfloat[Maximum memory consumption
  \label{fig:motivation-mem}]{%
        \includegraphics[width=0.5\linewidth]{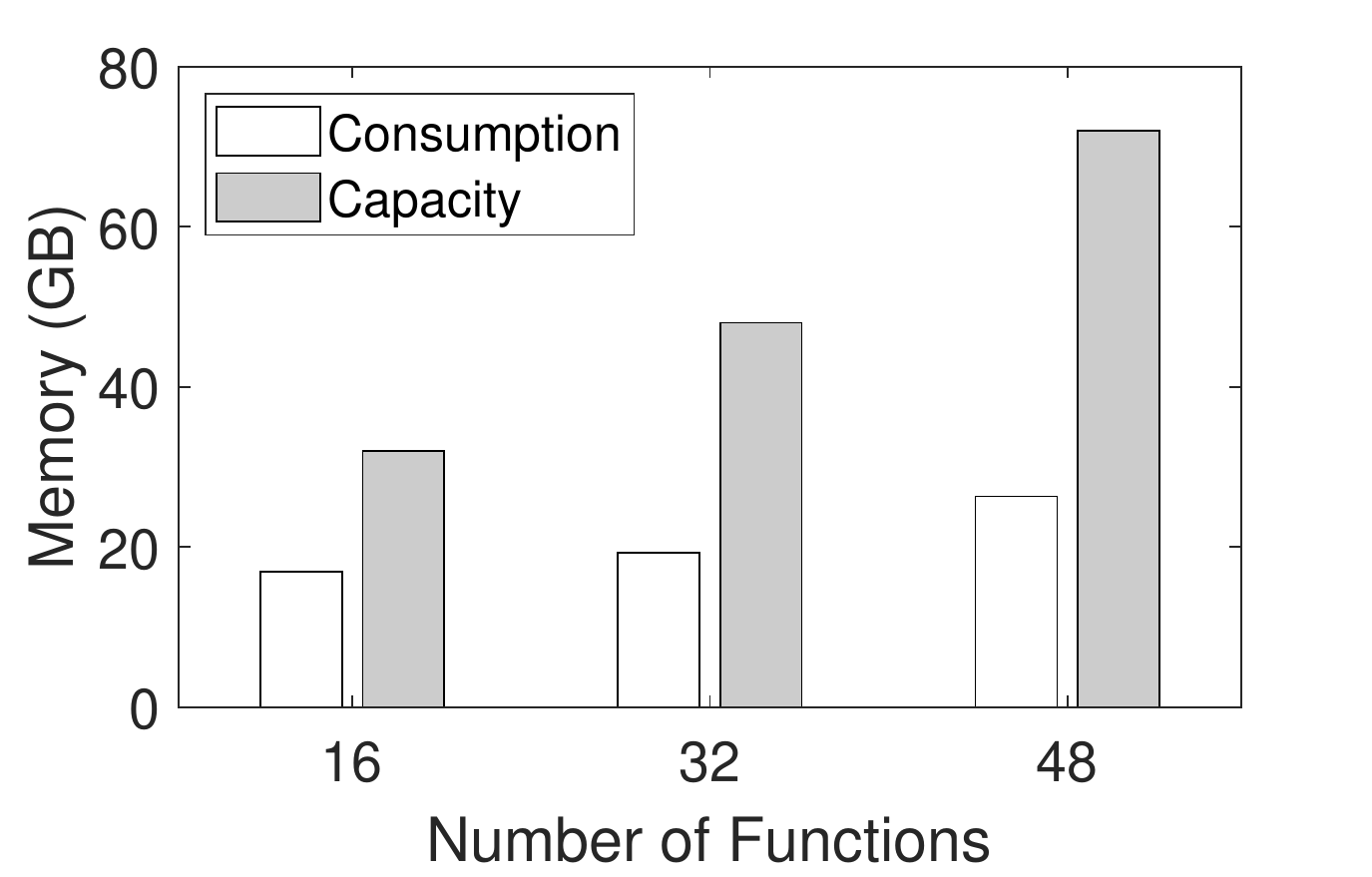}}
  \caption{Evaluation of an availability-aware approach in estimating the capacity of a FaaS platform and ensuring the SLA for response time}
  \label{fig:motivation} 
\end{figure}

We have run the simulation in three settings with 16, 32 and 48 functions and observed the effect of different hit rates over the function response times. The function idle times are set by solving (\ref{eq:che-hit-rate}) for the specific hit rate. This hit rate have also been used to estimate the memory capacity. The hit rate is related to the average  runtime memory consumption ($m$) as  $m= \sum_ih_i\theta_i$, where $\theta_i$ is the memory requirement of each functions \cite{fricker2012versatile}. We have used this value of $m$ as the memory capacity and compared it with the actual memory consumption value obtained from \textit{FaasSim}. The findings from the simulation are presented in Fig. \ref{fig:motivation}.

\subsection{Observations}
In Fig. \ref{fig:motivation-response}, we plot the response times of each of the 48 functions for different hit rates. We see that even with 95\% hit rate, there are response times that violate the SLA. However, for 95\% hit rate, more than half of the function response times are much lower than the SLA. This indicates that all the functions do not require the same hit rate to ensure the SLA. In Fig. \ref{fig:motivation-mem}, we present a comparison between the estimated memory capacity and maximum consumption for 95\% hit rate. Although the capacity notably increases with the number of functions, the consumption is less sensitive to it. This is because memory consumption is primarily dependent on the workload parameters. In addition, the consumption is not very high since most of the functions remain idle while resident in memory, which is not considered during the estimation. 


From these observations, it is clear that an availability-aware approach is not adequate for optimal capacity estimation that ensures the SLA for response time. Such an approach only considers the volume of cold starts, whereas we also need to consider its effect on the response time. For a particular workload, firstly, we should know the cold start probabilities of the functions for different idle times. Subsequently, depending on these probabilities and the severity of cold starts, we need to approximate the function response times. Thus, we need a performance model incorporating all these factors. The model will also help in fine-grained capacity estimation by providing the resource utilization estimates. In the following section we present our performance model.

\section{Modeling Cold Starts in FaaS}
\label{sec:model}

\subsection{Estimating Cold Start Probabilities}
\label{subsec:mm1-set-delay}

Unlike commercial FaaS platforms, open source platforms like OpenFaaS allow concurrent function execution in same container \cite{akkus2018sand}. We focus on this function concurrency approach. We propose to consider, from a modeling standpoint, the function as a server of a queueing model, representing the admission control buffer to the function, and the cold start delay as the initial setup time of the server before beginning service. The functions also have an idle time which is equivalent to the idle server waiting time before it is shut down. Considering these similarities, a cold start may be modeled as a M/M/1/setup/delayedoff model, which is a variant of M/M/k/setup class of models \cite{gandhi2013exact}. The M/M/k/setup models consider a setup cost, usually in the form of a time delay, when turning the server on. Its ``delayedoff'' variant considers an idle time before turning the server off.

\begin{figure}[t]
    \centering
    \includegraphics[width=0.8\linewidth]{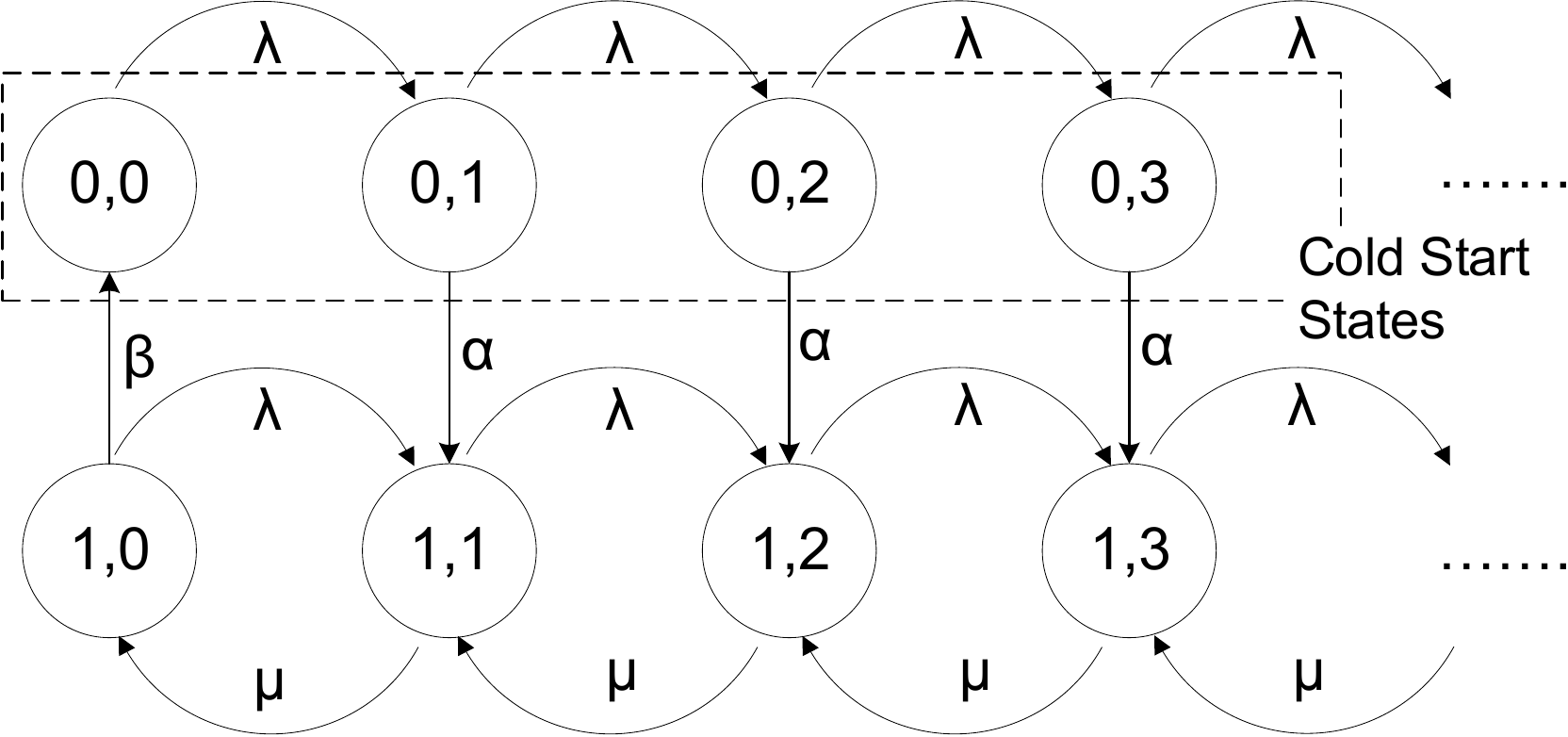}
    \caption{The M/M/1/setup/delayedoff model for a function}
    \label{fig:mm1-set-delay}
\end{figure}

Although in \cite{gandhi2013exact}, the exact solution is provided for an M/M/k/setup/delayedoff model, this applies when the number of servers is $k\geq2$. However, in our case we need to model each function separately. Thus, we have a function representing a single server, which can be either on or off. To get different performance indices for such a model, we may directly solve its underlying Continuous Time Markov Chain (CTMC). 

The CTMC transitions are presented in Fig. \ref{fig:mm1-set-delay}. Each CTMC state $(i,j)$ has two parameters: $i$ tracks whether the function resides in the memory or not $(i \in \{0,1\})$, while $j$ tracks the number of jobs $(j \in \mathbb{Z}^*)$ in the admission queue to enter service in the function. A transition from $(i, n)$ to $(i,n+1)$ occurs with rate $\lambda$, transition from $(i, n+1)$ to $(i,n)$ occurs with rate $\mu$, and transition from $(0, j)$ to $(1,j)$ occurs with rate $\alpha$. These rates describe the mean inter-arrival time $(\frac{1}{\lambda})$, mean service time $(\frac{1}{\mu})$ and mean cold start time $(\frac{1}{\alpha})$ respectively. There is a special transition from $(1,0)$ to the initial state $(0,0)$ with rate $\beta$, which describes the function idle time $(\frac{1}{\beta})$. In a CTMC, all holding times are considered to be exponentially distributed. However, in a real system the idle time of a function is set to a deterministic value. To address this issue, we can use the method of phases and make this transition Erlang-$k$ distributed with rate $k\beta$. To realize this, we introduce $k-1$ extra states between $(1,0)$ and $(0,0)$. The transitions between all these states occur with a rate $k\beta$. This keeps the mean identical to the original exponential, $\frac{1}{\beta}$, but reduces the variance by $k$ times. Thus, for large enough $k$, the transition will display a behavior close to deterministic. 

The effect of cold starts vary depending on the sequence of request arrivals. If a request arrives when the function is being loaded into memory due to a recent request, the response time of that request will be affected to some extent. The severity of the queueing overhead will depend on the residual cold start time of the  previous request. However, this does not need to be modeled explicitly thanks to the memoryless property of the exponential distribution. Considering this, as shown in Figure \ref{fig:mm1-set-delay}, it is clear that the cold start states are $(0,j), \forall j$. We can calculate the cold start probability of the functions from the stationary distribution ($\pi$) of their CTMC. Indicate with $\pi_{i,j}$ the probability of state $(i,j)$, then the cold start probability is defined as $\sum_j \pi_{0,j}$. We can get the stationary distribution by solving the CTMC. This can be done efficiently using the matrix-analytic method, since the CTMC sparsity structure makes it equivalent to a M/G/1-type process \cite{LatoucheR99}. The latter is analyzed using the implementation in the MAMSolver~\cite{riska2002mamsolver,riska2007etaqa}.


\subsection{Predicting Response Time}
\label{subsec:lqn-model}

Solving the CTMC we can get the cold start probabilities for each of the functions. However, our eventual goal is to predict the response time of each functions considering the cold starts. For that, beside the cold start probabilities, we need a performance model of the functions, typically running in containers, contending for the CPU. Each of these functions contend for CPU times to execute two types of jobs, the regular tasks when the function is warm and service restarting when the function is cold. To ensure scalability of the model analysis \cite{tribastone2010performance}, we use LQNs as reference modeling formalism.

\begin{figure}[t]
    \centering
    \includegraphics[width=0.7\linewidth]{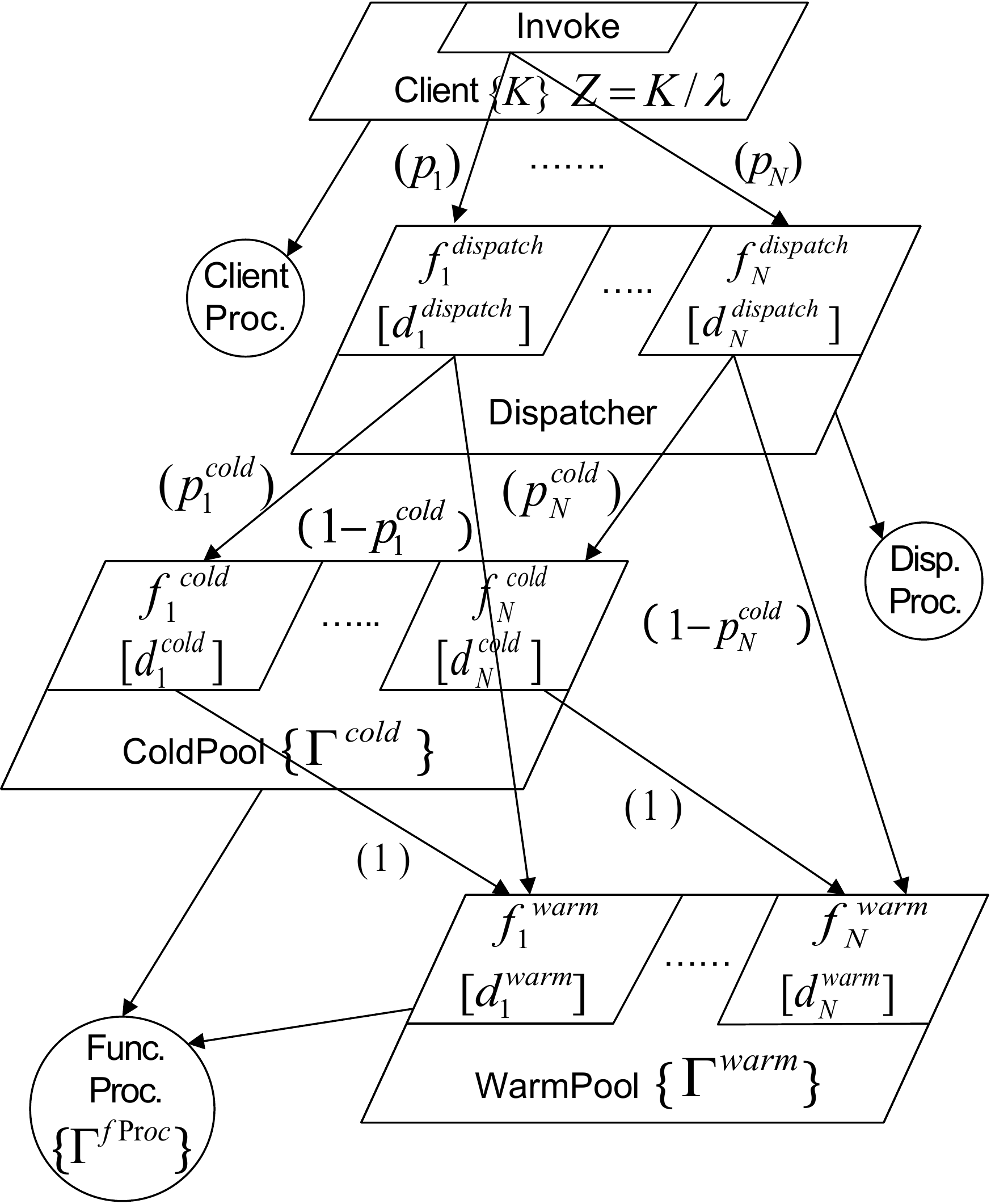}
    \caption{The LQN model for a FaaS platform}
    \label{fig:faas-lqn}
\end{figure}


The proposed LQN model\footnote{For details about the notation, please see the LQN user manual available at \url{http://www.sce.carleton.ca/rads/lqns/LQNSUserMan-jan13.pdf}} is presented in Fig. \ref{fig:faas-lqn}. The model has two main building blocks - the tasks and the processors. In LQN models, tasks translate into different system resources, usually the software resources. They carry out different operations which are defined by their entries. The tasks are executed on the processors, which represent the physical entity, like the CPU, that carries out the physical executions. Although each of the functions is a software resource, we defined them by the entries rather than the tasks. The reason behind this choice is twofold. Firstly, it makes the LQN model more compact and manageable. Secondly, it reduces the model solving delay as the number of function increases.

Since each function has two types of jobs, we use two tasks, \textit{ColdPool} and \textit{WarmPool}. The entries in the \textit{ColdPool} define the cold jobs for all the functions. Similarly, the entries in the \textit{WarmPool} define the warm jobs. Since every cold job is followed by a warm job, there is a call from the cold entries to the warm entries. The proportion of cold and warm jobs is controlled by the \textit{Dispatcher} task based on the cold start probabilities. This is done by setting the cold start probability of each function to the call mean value from its \textit{Dispatcher} entry to the \textit{ColdPool} entry. The percentage of calls to each functions, based on their popularity, is modeled using the reference task \textit{Client} by setting the percentage value as the call mean from the \textit{Client} entry to the \textit{Dispatcher} entry. 

The LQN model requires two parameters, namely the service demands of the activities and the multiplicities of the modeling constructs. The service demand for a job is the total service time across all visits when there is no resource contention. Each of the functions has different service demands for its cold and warm jobs. These values should be set in the activities of the corresponding entries. The service demand can be estimated using state of the art techniques based on utilization or response time \cite{gias2019atom}. The multiplicities translate into different system entities depending on the modeling constructs. The multiplicity of the reference task indicates the number of clients present in the system, considering the system as a closed network \cite{harchol2013performance}. However, we can also consider the system as open like our \textit{FaasSim} simulator. To do so, we have adapted the think time $Z$ as $K/\lambda$, where $K$ and $\lambda$ represents the total number of clients and the open arrival rate~\cite{shousha1998applying}.

The multiplicities of the processors indicate the number of available CPU cores. Since we do not consider the \textit{Dispatcher} a bottleneck, we assume it executes separately from the functions, on a single CPU core. The multiplicities of the \textit{ColdPool} and \textit{WarmPool} indicate the number of process threads available for the function containers. Container platforms like Docker allow this on a container basis, which means that we can put a limit on how many threads a container can create\footnote{Such limits are put to prevent unnecessary thread creation causing memory leaks. However, the limits are never too small to affect the concurrency.}. However, in LQNs entries do not have a multiplicity property, which we are using to model the functions. Thus, in the model, we consider that the functions share two thread pools for cold and warm jobs. This assumption does not significantly affect the performance estimates if the number of threads, in both pools, are sufficiently large to start processing a job immediately.


\subsection{Model Validation}
\label{subsec:model-validation}

\begin{table}[t]
\centering
\caption{Simulation parameters for model validation}
\label{tbl:model-validation-param}
\begin{tabular}{|lll|}
\hline
$N$ & Number of functions & 16, 32, 64, 96, 128 \\
$\eta$ & Zipf parameter & 0.6, 1.0, 1.4 \\
$\lambda$ & Arrival rate & 0.2, 0.5, 0.8 \\
$\mu$ & Service rate & [1, 2] \\
$\alpha$ & Cold start rate & [0.037, 0.5] \\
$\beta$ & Idle lifetime rate & [0.00083, 0.00556] \\ \hline
\end{tabular}
\end{table}

We have used the LINE performance modeling tool \cite{casale2019automated} to build our model and validated it with the \textit{FaasSim} simulator. The simulation parameters for the experiments are presented in Table \ref{tbl:model-validation-param}. We have considered more large-scale settings compared to Section \ref{sec:analogy} that includes up to 128 functions. 
We have also considered different popularity parameters which are common in cache based studies \cite{casale2018analyzing}. We have considered each of the combinations of number of functions ($N$), Zipf parameters($\eta$) and arrival rates ($\lambda$) from the table. For each of those combinations we have generated 30 models. In each of the models, we have chosen the service ($\mu$), cold start ($\alpha$) and idle lifetime ($\beta$) rates for the functions randomly from the given range. The range for service and cold start rates are same as Section \ref{sec:analogy}. The idle lifetime rates are chosen from \cite{mikhail2019comparison} such that it can trigger a cold start.

\begin{table}[t]
\centering
\caption{Percent error in estimating the response time of each function--Across all the Zipf parameters in Table \ref{tbl:model-validation-param}}
\label{tbl:model-validation-results}
\resizebox{\columnwidth}{!}{%
\begin{tabular}{|c|c|c|c|c|c|c|c|c|c|}
\hline
\multirow{2}{*}{\textbf{N}} & \multicolumn{3}{c|}{\textbf{$\lambda$ = 0.2}} & \multicolumn{3}{c|}{\textbf{$\lambda$ = 0.5}} & \multicolumn{3}{c|}{\textbf{$\lambda$ = 0.8}} \\ \cline{2-10} 
 & \textbf{avg} & \textbf{95p} & \textbf{max} & \textbf{avg} & \textbf{95p} & \textbf{max} & \textbf{avg} & \textbf{95p} & \textbf{max} \\ \hline
\textbf{16} & 1.24 & \textbf{2.08} & \textbf{2.38} & 0.87 & 1.57 & 2.2 & 0.91 & 1.48 & 1.65 \\ \hline
\textbf{32} & 1.25 & 2.01 & 2.3 & 1.17 & 1.87 & 2.21 & 1.15 & 1.75 & 2.16 \\ \hline
\textbf{64} & 0.85 & 1.7 & 2.06 & 1.21 & 1.87 & 2.15 & 1.27 & 1.72 & 2.09 \\ \hline
\textbf{96} & 0.82 & 1.36 & 1.98 & 1.11 & 1.57 & 1.83 & 1.25 & 1.7 & 2.05 \\ \hline
\textbf{128} & 0.92 & 1.31 & 1.71 & 1.11 & 1.51 & 1.95 & \textbf{1.29} & 1.7 & 2.1 \\ \hline
\end{tabular}%
}
\end{table}

Since cold start affects the response time, we are concerned that how accurately our model captures that affect and estimate the response time of each functions. Thus, we have considered the percent error in estimating the function response times. We present the results in Table \ref{tbl:model-validation-results}. From the table, we  see that the increase of the number of functions has a negligible effect over the error. The maximum error across all the parameters is 2.38\%. The maximum average error and $95^{th}$ percentile of the error is 1.29\% and 2.08\% respectively. Such errors are not significant and thus we conclude that the LQN model can accurately estimate the response times for each functions considering the cold starts. 
 
We leverage this model in our capacity estimation method COCOA, which we present in the following section.

\section{Cold Start Aware Capacity Planning}
\label{sec:refine}

\subsection{Overview}
\label{subsec:cocoa-approach}

\begin{figure}[t]
    \centering
    \includegraphics[width=0.85\linewidth]{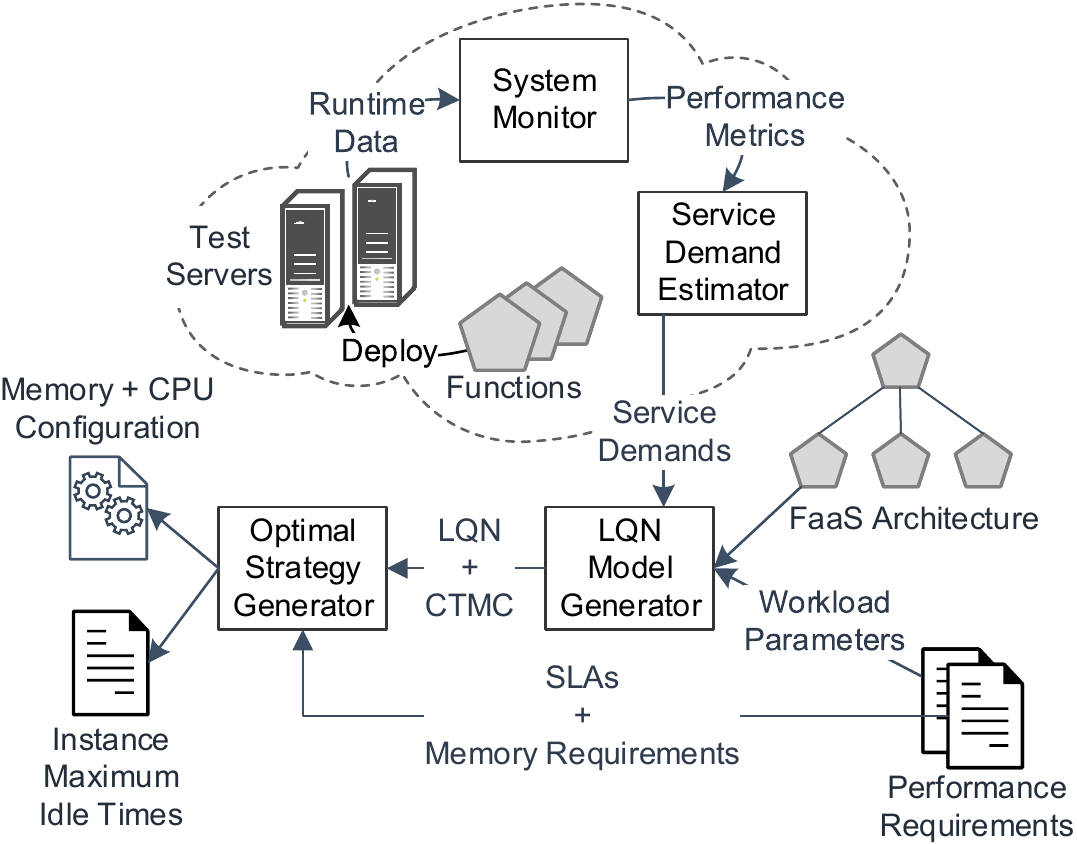}
    \caption{An overview of the COCOA approach}
    \label{fig:cocoa-approach}
\end{figure}

COCOA provides resource allocation decisions for a FaaS platform in terms of its memory and CPU configurations. It also provides the function idle times that can ensure the SLA given the suggested configuration is applied. These idle times allows to control the magnitude of cold starts, which in turn aids in governing the response times. The configurations can be applied both on the hardware level or on the software level. This means that the memory and CPU constraints can be applied on the physical server or the container platform like Docker. COCOA has multiple components each focusing on a particular tasks and expects different inputs. These components, their expected inputs and outputs are illustrated in Figure \ref{fig:cocoa-approach}.

Since COCOA is a model-based approach, we need a set of parameters to instantiate its model. Firstly, we need the service demands of each functions. These can be estimated for each functions individually when they are being developed in a test environment. The workload parameters like the arrival rate and function popularity can be estimated from performance requirements or historical data. Once the parameters are estimated they are passed to the component LQN Model Generator. The Model Generator also requires the architecture of the FaaS platform, particularly containing the information about how the functions communicate.

Based on these inputs, the CTMC and LQN model is generated and forwarded to the next component, the Optimal Strategy Generator. It utilizes the models to provide memory and CPU configurations. It needs the SLAs for function response times and both the memory requirements when they are idle and in execution. The generator then searches for the idle times, under different CPU configurations, that do not violate the SLA with minimal memory consumption. These idle times are used to estimate the maximum memory consumption, based on which the memory capacity is suggested. 


\subsection{Problem Statement}
\label{subsec:problem}

We consider a system of $N$ functions. These functions are executed on a multi-core CPU with $C$ cores. Each of the functions has two different memory usage, one is while in execution ($\theta_i^{\text{on}}$) and the other is while being idle ($\theta_i^{\text{off}}$) 
Considering a request for a function $f_i$, if the function is not in the memory, a cold start occurs. Due to this cold start, a request experiences an extra delay. This extra delay is incurred to load the function in the memory. When a function is loaded in the memory, it is associated with a timeout value $T_i$. While the function is still in the memory, for each new request, the timeout is reset to the original value. A function is removed from the memory if it reaches the timeout limit.

We focus on two specific costs, the cost of CPU and cost of the memory. We define the per unit CPU and memory cost as $\tau_c$ and $\tau_m$ respectively. Thus, the cost for the CPU will be $B= \tau_c C$. 
The memory cost is calculated based on maximum memory consumption. To estimate this, we incorporate the idea of a different memory usage, when the function is idle, with the estimator for TTL cache \cite{fricker2012versatile}. Based on this, given the function CPU utilization is $\rho_i$, the average memory consumption ($m$) may be estimated 
as $m= \sum_i h_i(\rho_i\theta_i^{\text{on}} + (1-\rho_i)\theta_i^{\text{off}})$.

The system's memory capacity should be adequate when there is a spike in memory consumption. This occurs when there is a surge in requests within a short period. This increases the memory consumption because more functions starts execution, for which the memory requirement is much higher than being idle. It is sufficient to consider this increase in memory consumption by the functions in execution. Considering the memory consumption by the functions in execution is $U$, the expectation is defined as $E[U]=\sum_i h_i \rho_i \theta_i^{\text{on}}$. We can approximate the maximum consumption as $v=\kappa E[U]$. The value of $\kappa$ is calculated, using Markov's inequality, \cite{nelson2013probability} such that the upper-bound of $P(U \geq v)$ is a negligible value $\epsilon$. Based on this, we define the approximation for maximum memory consumption ($m^{\max}$) as $m^{max}= \sum_i h_i(\kappa\rho_i\theta_i^{\text{on}} + (1-\rho_i)\theta_i^{\text{off}})$, so that the memory cost may be defined as $A=\tau_m m^{\max}$.

Considering these cost functions $A$ and $B$, our objective function, $z$, is defined in (\ref{eq:objective}). Here, our goal is to find $T$, a vector including the idle times $T_i$ of all the functions, and $C$, the number of CPU cores, that minimizes a weighted sum of normalized memory and CPU cost. 

\begin{IEEEeqnarray}{CC}
z= \min_{(T,C)}  \omega_A \hat{A} + \omega_B \hat{B}\label{eq:objective}
\\
\text{subject to:}\nonumber\\
C \leq C^{\max} \label{eq:constraint1}\\
W_i(T, C) \leq W^{*}, \forall i \label{eq:constraint2} \\
\qquad T \in \mathbb{R}_{+}^N, C \in \mathbb{Z}_{+} \nonumber 
\end{IEEEeqnarray}

The constraints for the objective function are provided in \eqref{eq:constraint1} and \eqref{eq:constraint2}. The first constraint in \eqref{eq:constraint1} is regarding the maximum number of allowed CPU cores. This applies when the CPU constraint is imposed on the software level and the total physical CPU capacity is not accessible. The second constraint in \eqref{eq:constraint2} addresses the SLA for response time. Here, $W_i$ is a function of $T$ and $C$ which returns the response time for a platform function $f_i$. This response time should be less than the limit $W^*$ mentioned in the SLA.

\subsection{Optimal Strategy Generation}
\label{subsec:capacity}

Using the objective function in (\ref{eq:objective}), COCOA provides an optimal strategy including the memory and CPU capacity and the function idle times ($T$). It starts searching for an optimal strategy with an initial instance of $T$. This is obtained  by a characteristic time approximation technique for CDN cache \cite{che2002hierarchical}. It requires to solve $m=\sum_ih_i$, where $h_i$ is defined in \eqref{eq:che-hit-rate}, for a particular value of $m$. However, since we have a large pool of functions, as suggested in \cite{martina2014unified},  we have estimated a single value $T^*$ for all the functions instead of approximating $T_i \in T, \forall_i$. Thus, here we have used a second definition of $h_i$, replacing $T_i$ with $T^*$ in \eqref{eq:che-hit-rate}.


After the initialization, COCOA fine-tunes the idle times, such that the function response times are just under the SLA limit, to ensure minimal memory consumption. For this, it solves the LQN model in iteration, upon adjusting the idle times, to observe its effect on the response time. The idle times are adjusted using the concept of binary search. It starts with an initial searching interval, $(0 \enskip T^*]$, for each $T_i$ and reduces the length of the interval by half on each iteration. The endpoints of the intervals are adjusted depending on whether the response time constraint is satisfied or not. The value of $T_i$ is updated with the midpoint of the searching interval. For this process to work, the initial value, $T^*$, should be sufficiently large so that there is no cold starts and thus the response time is not affected. For this purpose, we have solved $m=\sum_ih_i$ by setting $m$ to a value close to $N$.

COCOA runs this fine-tuning process for different CPU configurations (CPU cores). Although the number of CPU cores can be any integer, practically we only need to consider some common options, like multiples of 2 with 32 as the limit. This accelerates the analysis process. In addition, for each configuration, this process is run in parallel, making it even faster. For each run, if a $T$ is found, that does not violate the SLA, it is considered as a candidate solution. After completing the process, the optimal solution is selected by comparing the memory and CPU cost. Its corresponding CPU configuration and idle times are suggested just the same. However, the memory capacity is suggested by considering the value $\min(m^{\max}, \sum_i\theta_i^{\text{on}})$ as an upper-bound and calculating the aggregated size of required number of RAM modules\footnote{We have considered that each of the memory module is 8GB but this is configurable depending on the availability of RAM modules.}.

\section{Evaluation}
\label{sec:result}


\begin{figure}[t]
    \centering
    \includegraphics[width=0.65\linewidth]{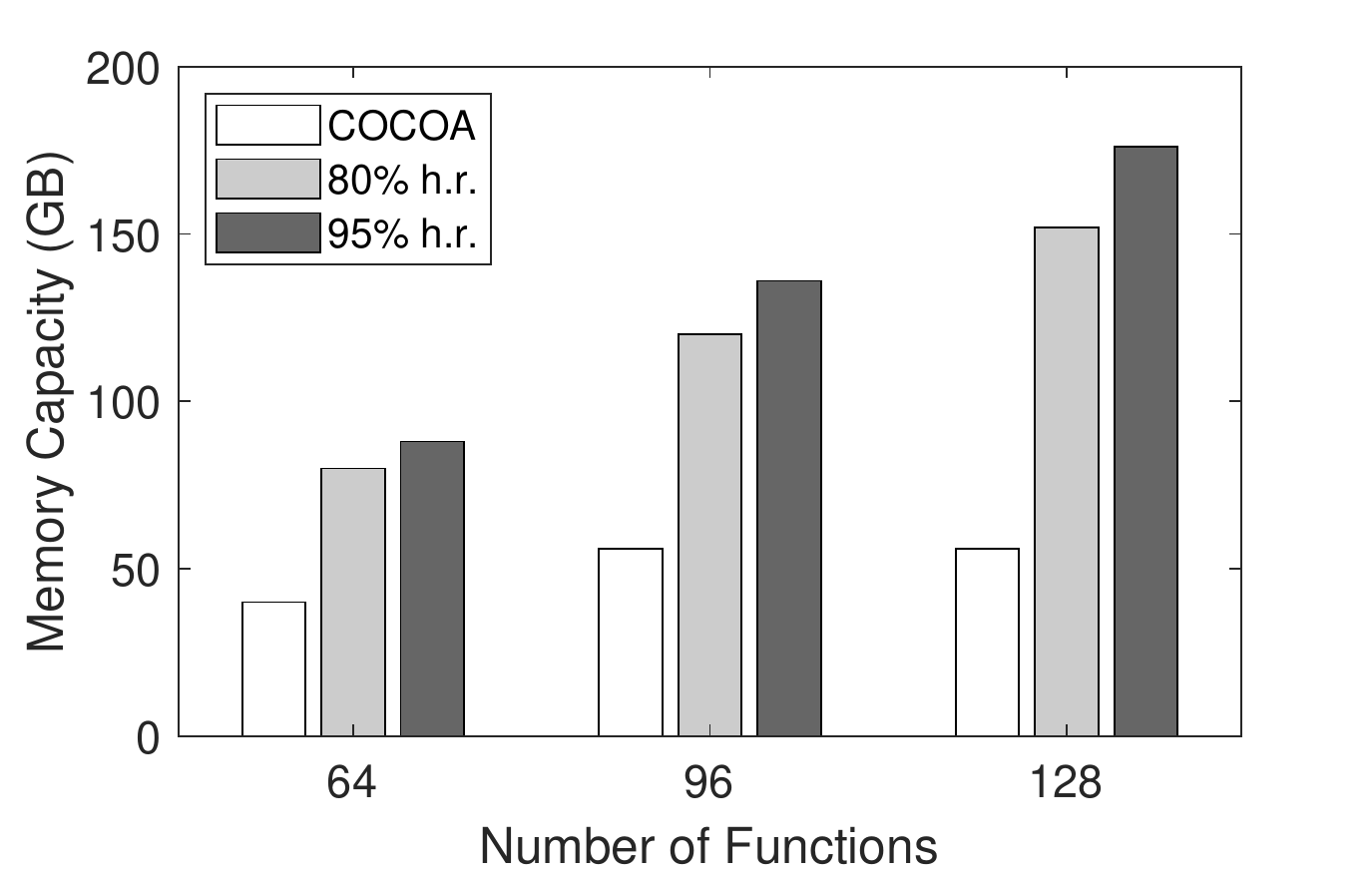}
    \caption{Comparing the memory capacity predicted by COCOA and the availability-ware approaches with $\lambda=0.5$ and $\eta=1.0$}
    \label{fig:eval-mem}
\end{figure}

\subsection{Experimental Setup}

We have evaluated COCOA using the \textit{FaasSim} simulator, considering the parameters from Table \ref{tbl:model-validation-param} with 64, 96 and 128 functions. However, the idle times ($\frac{1}{\beta}$) from Table \ref{tbl:model-validation-param} are not used. Instead, these have been estimated with COCOA such that the response times constraints are satisfied with minimal memory and CPU requirement. We have set the functions memory requirement following the limits in AWS Lambda \cite{AWSLambdaLimit}. The percentage (0-1) of idle function memory consumption is considered to be log-normally distributed with a desired mean of 0.2. From the experiments, we aim to answer the following research questions -

\begin{itemize}
	\item \textbf{RQ1: } Can COCOA reduce memory over-provisioning compared to availability-aware approaches?
	\item \textbf{RQ2: } Can COCOA predict the required memory capacity that meets the maximum demand?
	\item \textbf{RQ3: } Can COCOA predict the memory and CPU capacity to satisfy  the SLA for response time?
\end{itemize}

\begin{table*}[t]
\centering
\caption{Comparing the predicted memory capacity of different approaches - averaged across the Zipf parameters from Table \ref{tbl:model-validation-param}}
\label{tbl:cocoa-evaluation}
\begin{tabular}{|c|c|c|c|c|c|c|c|c|c|}
\hline
\multirow{2}{*}{\textbf{N}} & \multicolumn{3}{c|}{\textbf{$\lambda$ = 0.2}} & \multicolumn{3}{c|}{\textbf{$\lambda$ = 0.5}} & \multicolumn{3}{c|}{\textbf{$\lambda$ = 0.8}} \\ \cline{2-10} 
 & \textbf{COCOA} & \textbf{80\% h.r.} & \textbf{95\% h.r.} & \textbf{COCOA} & \textbf{80\% h.r.} & \textbf{95\% h.r.} & \textbf{COCOA} & \textbf{80\% h.r.} & \textbf{95\% h.r.} \\ \hline
\textbf{64} & 32 & 88 & 104 & 45.3 & 80 & 93.3 & 42.7 & 85.3 & 101.3 \\ \hline
\textbf{96} & 40 & 128 & 152 & 58.7 & 125.3 & 146.7 & 48 & 120 & 141.3 \\ \hline
\textbf{128} & 48 & 157.3 & 184 & 58.7 & 157.3 & 184 & 64 & 168 & 197.3 \\ \hline
\end{tabular}%
\vspace{-2em}
\end{table*}

\subsection{Results}
\label{subsec:results}

To answer \textbf{RQ1}, we have compared COCOA and the availability-aware approach with two hit rates, 0.8 and 0.95. We present the result for a single experiment in Fig. \ref{fig:eval-mem}. We can see that the required memory capacity estimated by COCOA is much lower than the other two approaches. The results for all the parameters are presented in Table \ref{tbl:cocoa-evaluation}. In all the cases, the estimates from COCOA is much less compared to the other two approaches. Considering the 95\% hit rate, the capacity estimated by COCOA is 51-74\% less. The reason is easy to understand - COCOA can take ``well-informed'' decisions by leveraging its performance model, which is not possible for the availability-aware approaches.

From Table \ref{tbl:cocoa-evaluation}, we see that, in two cases, COCOA predicts a higher capacity for $\lambda=0.5$ than $\lambda=0.8$, which is counter-intuitive. This is because we have used a different upper-bound of $P(U>v)$ to estimate $\kappa$ for $\lambda=0.8$. For $\lambda=0.8$ it is $0.1$ but for $\lambda=0.2$ and $0.5$ it is $0.05$. The reason is, for high arrival rates, the spike in memory consumption from the expectation is less than low arrival rates. Here, $\kappa$ is the coefficient to represent this extent. A larger upper-bound of $P(U \geq v)$ will result in a smaller value of $\kappa$. So for higher arrival rates, to reduce over-provisioning, $\kappa$ should be approximated with a larger upper-bound of $P(U \geq v)$.

\begin{figure}[t]
    \centering
      \subfloat[Avg. Memory Consumption \label{subfig:avg-mem}]{%
      \includegraphics[width=0.5\linewidth]{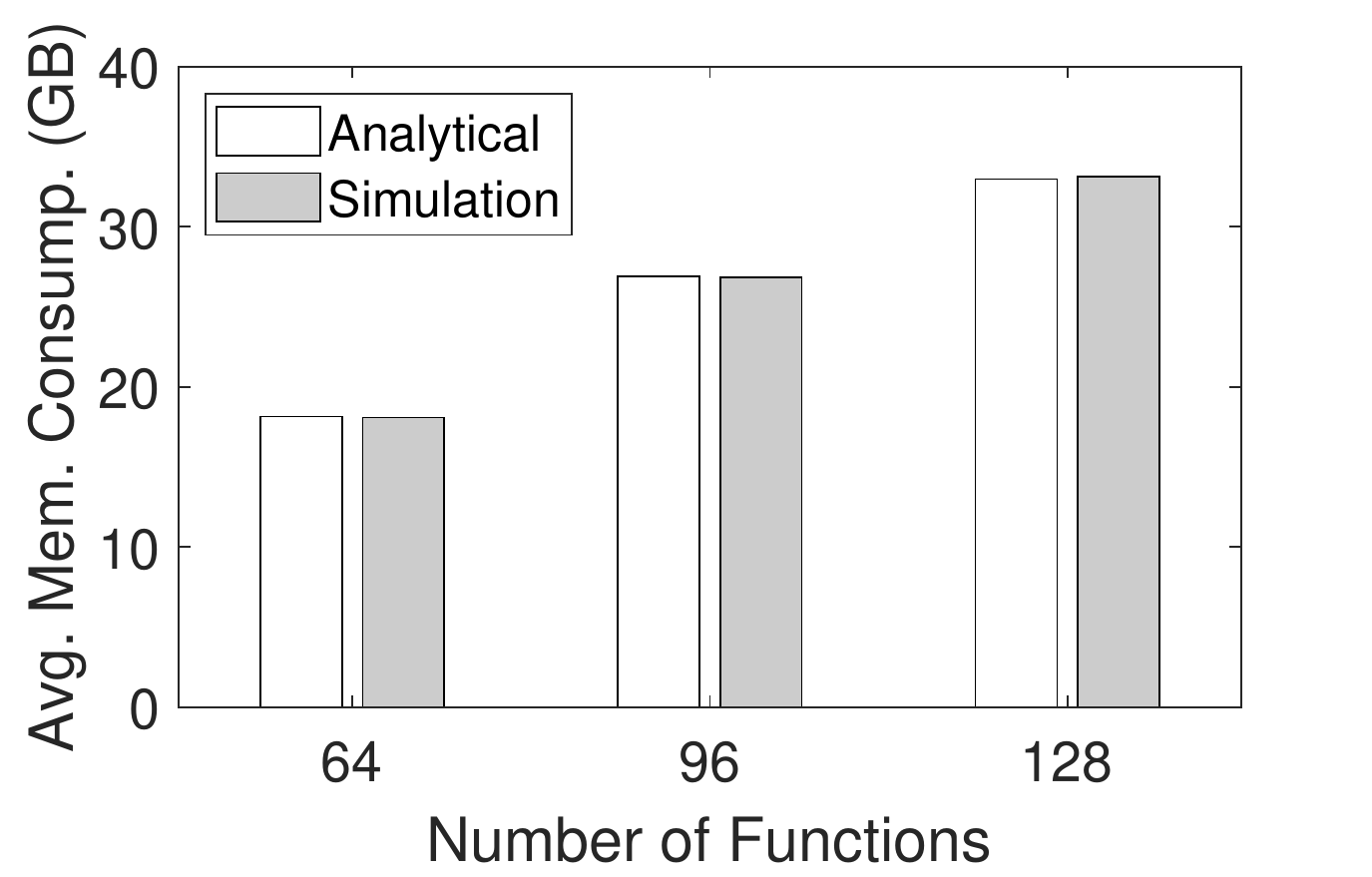}}
    \hfill
  \subfloat[Capacity vs. Consumption \label{subfig:capacity-consumption}]{%
        \includegraphics[width=0.5\linewidth]{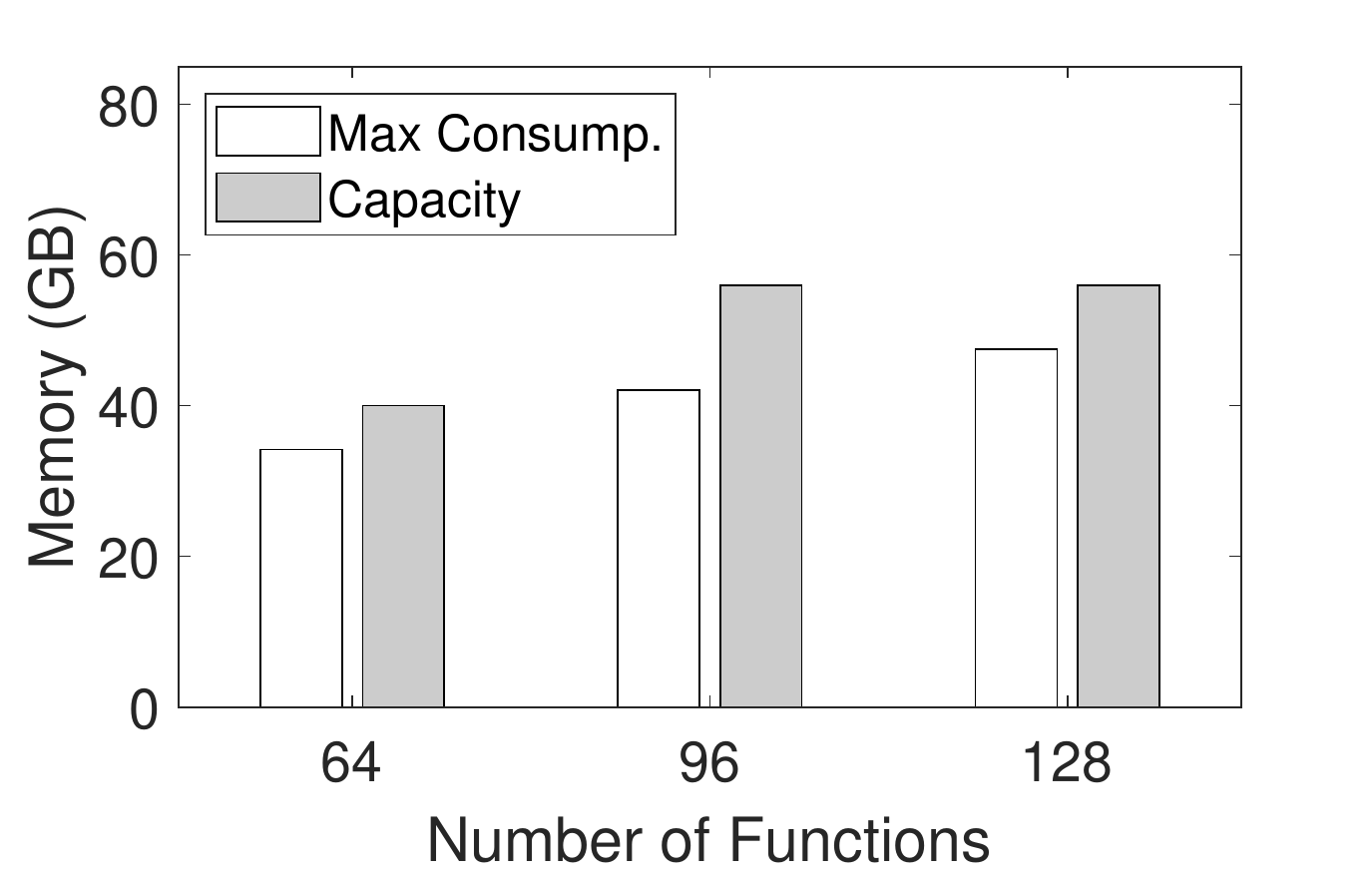}}
    \caption{Comparing the runtime memory consumption obtained from the analytical approximation and simulation for $\lambda=0.5$ and $\eta=1.0$}
    \label{fig:analytical-simulation-mem}
    \vspace{-1em}
\end{figure}

\begin{figure}[t] 
    \centering
  \subfloat[64 Functions \label{subfig:idle-mem-64f}]{%
      \includegraphics[width=0.5\linewidth]{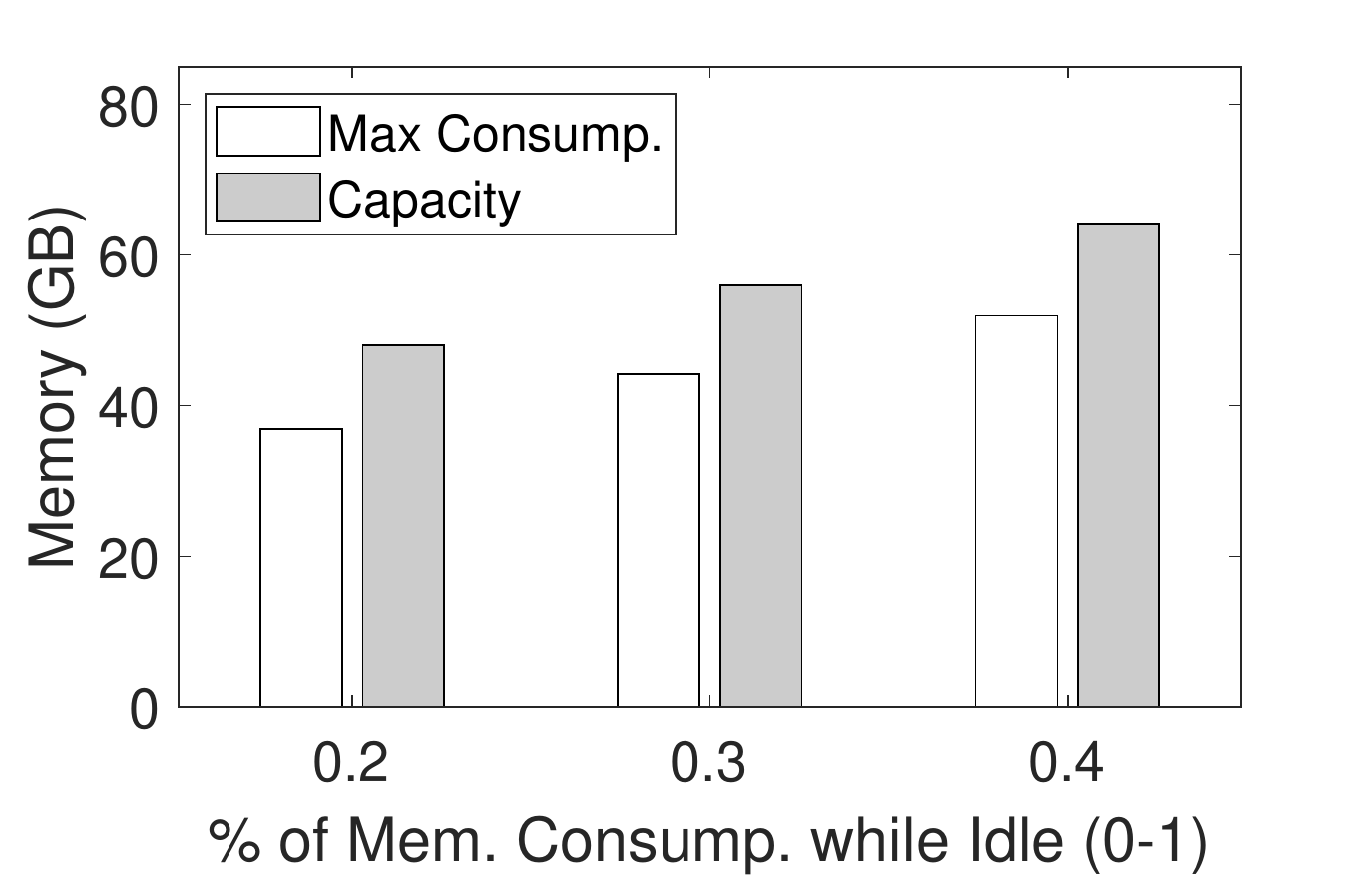}}
    \hfill
  \subfloat[128 Functions \label{subfig:idle-mem-128f}]{%
        \includegraphics[width=0.5\linewidth]{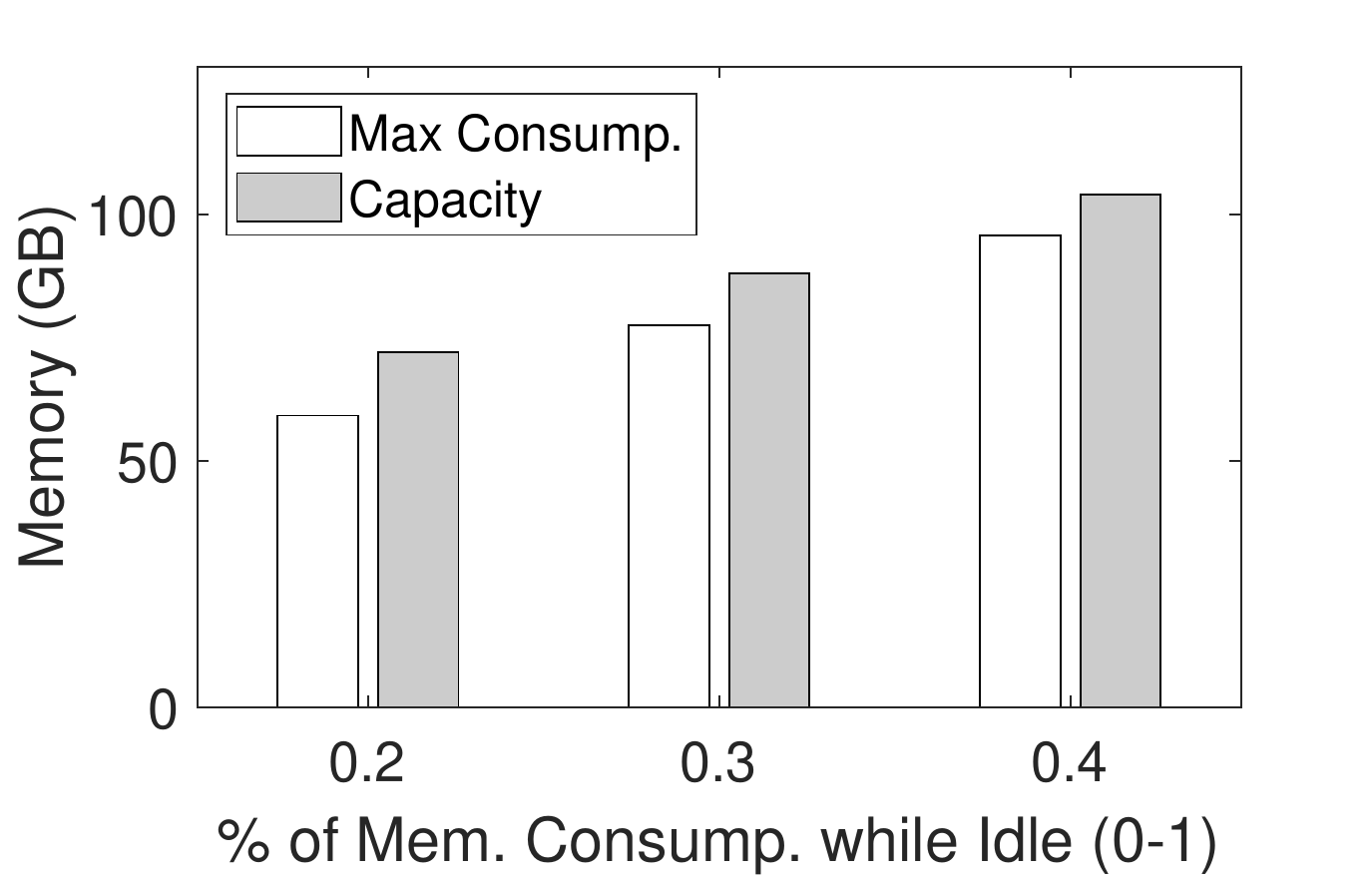}}
  \caption{Illustrating that COCOA can meet the maximum memory demand for different percentage of memory consumption while the functions are idle}
  \label{fig:sensitivity-idle-mem} 
\end{figure}

\begin{figure}[t] 
    \centering
  \subfloat[Response time\label{fig:eval-response}]{%
        \includegraphics[width=0.5\linewidth]{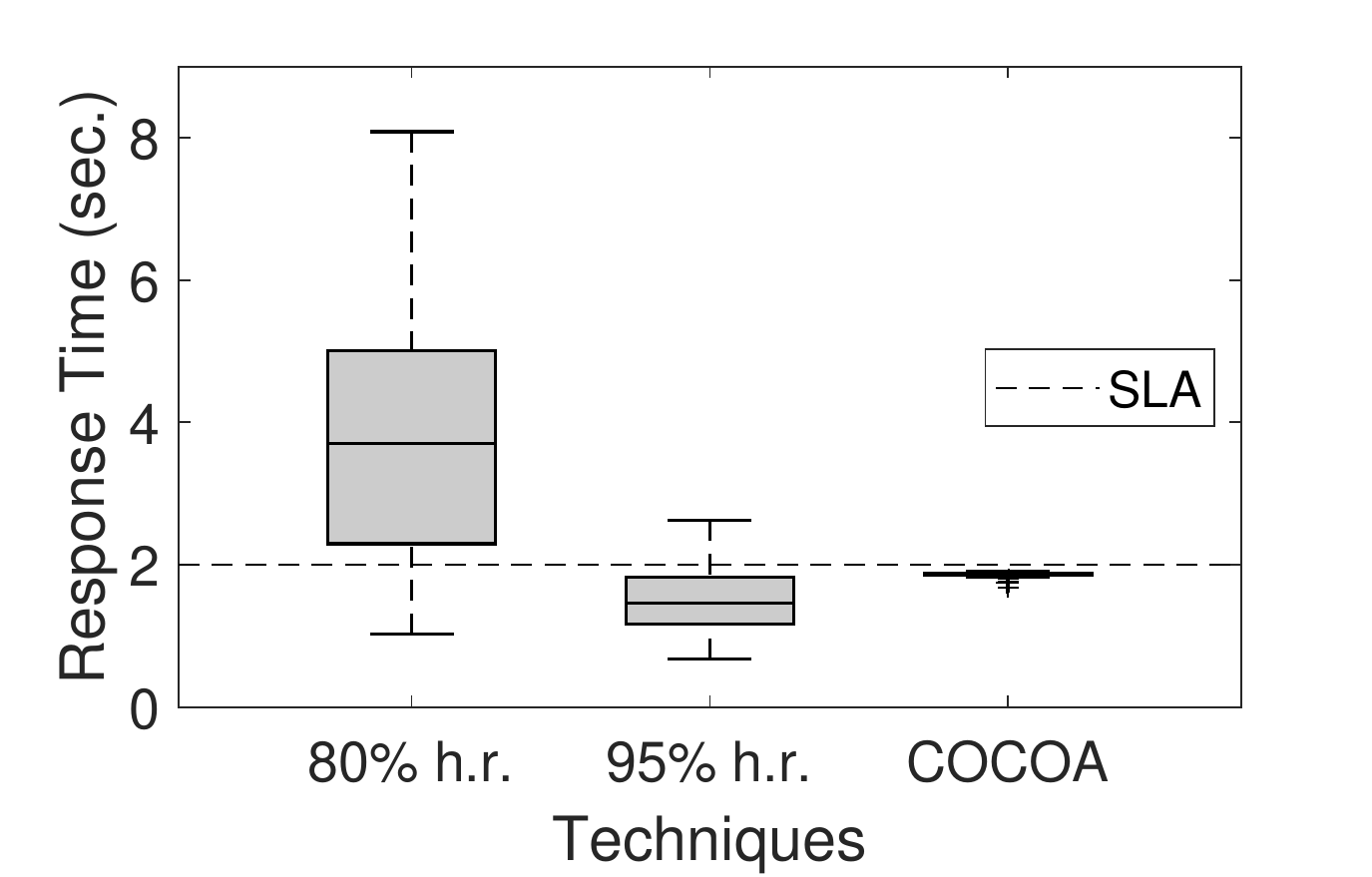}}
  \subfloat[Hit rate \label{fig:eval-hit}]{%
        \includegraphics[width=0.5\linewidth]{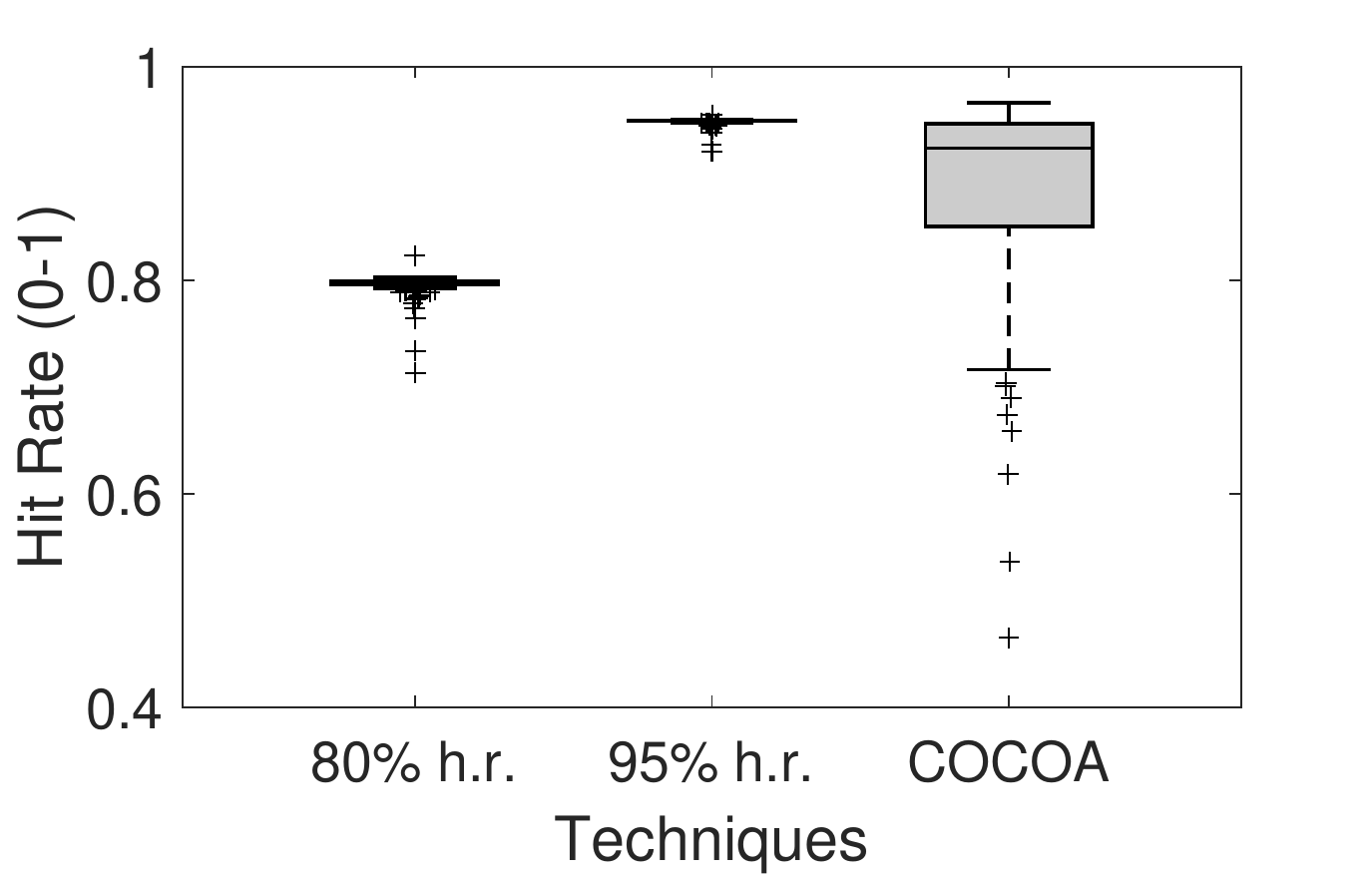}}
  \caption{Function response times and hit rates when the SLA is 2 seconds}
  \label{fig:evaluation-response}
\end{figure}

\begin{figure}[t] 
  \subfloat[Response time\label{fig:eval-response1}]{%
        \includegraphics[width=0.5\linewidth]{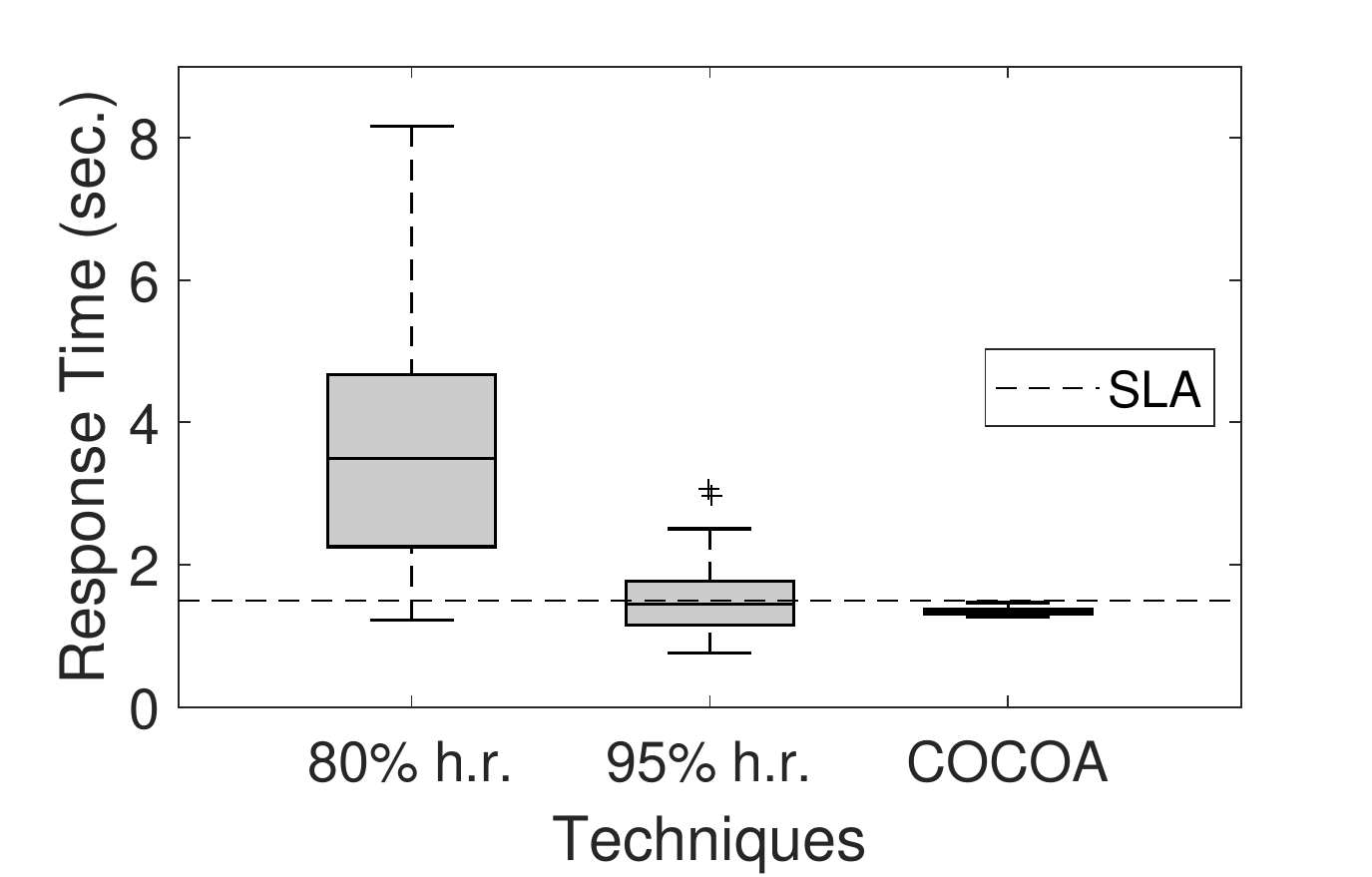}}
  \subfloat[Hit rate \label{fig:eval-hit1}]{%
        \includegraphics[width=0.5\linewidth]{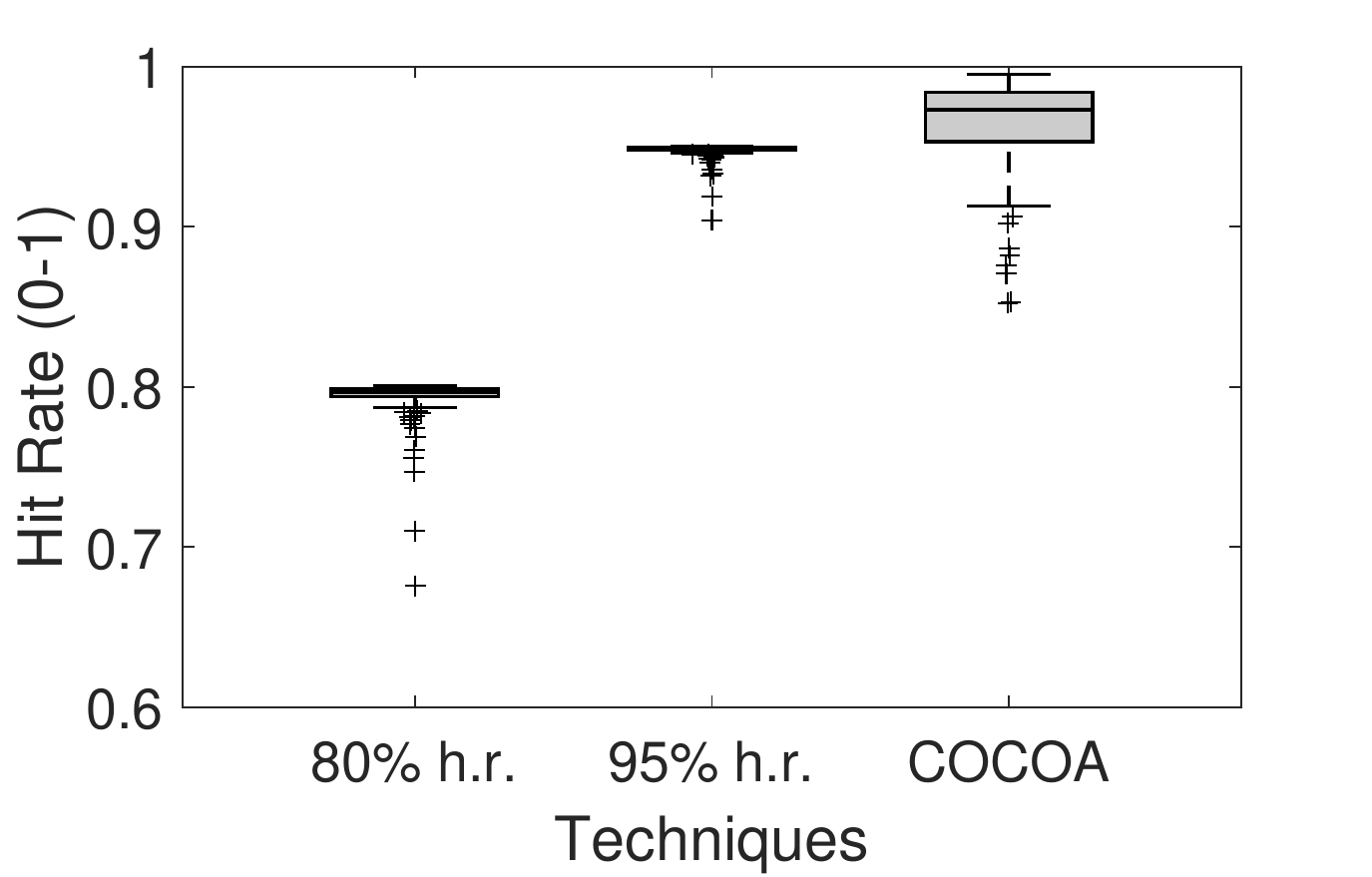}}
  \caption{Function response times and hit rates when the SLA is 1.5 seconds}
  \label{fig:evaluation1} 
\end{figure}

To answer \textbf{RQ2}, we have compared the memory consumption values from the simulation with the values from COCOA. From Fig. \ref{subfig:avg-mem}, we can see that the average memory consumption values from the simulation agrees with analytical approximation from COCOA. From Fig. \ref{subfig:capacity-consumption}, we see that the memory capacity also meets the maximum demand. From all the experiments, we observe only 5 cases where there is a memory deficit greater than 0.5 GB with a maximum value of 3.2 GB. We have also done a sensitivity analysis changing the desired mean of percentage of memory consumption by idle functions. We used two settings with 64 and 128 functions with $\lambda=0.8$ and $\eta=1.0$. As seen from Fig. \ref{fig:sensitivity-idle-mem}, in both cases, COCOA can satisfy the maximum demand.

To answer \textbf{RQ3}, we have investigated the response time of each of the functions. We have seen that across all the parameters, COCOA can ensure the SLA for response time. We present the response time of each function, for a single experiment, in Fig. \ref{fig:evaluation-response}. The SLA in this case is 2 seconds and COCOA satisfies it for all the functions with hardly any variance. On the other hand, even 95\% hit rate has violations. The violations are even more, 45\% or 57 out of 128 functions, when the SLA is 1.5 seconds.

To illustrate how COCOA ensures the SLA without over-provisioning, we have investigated the hit rates of each function obtained from the simulation. In Fig. \ref{fig:eval-hit} and \ref{fig:eval-hit1}, we have plotted the hit rates, which correspond the experiments from Fig. \ref{fig:eval-response} and \ref{fig:eval-response1}. As expected, we see that the hit rates are fixed for 80\% and 95\% hit rates. However, COCOA adjusts the idle times of the functions such that the hit rates are just sufficient to satisfy the SLA. This reduces the memory consumption when the functions are idle and thus COCOA suggests a much lower memory capacity. For a 2 seconds SLA, the lowest hit rate a function has is 46\%. However, COCOA can also increase the hit rates, if required, as seen in Fig. \ref{fig:eval-hit1}.  Here the hit rates for some functions are even higher than 95\% to satisfy a stricter SLA of 1.5 seconds.



\section{Related Work}
\label{sec:related}
FaaS platforms, leveraging serverless computing,  has gained the attention of many researchers. Here, we particularly focus on the works involving cost, resource management or cold stars as such works are more relevant in our context. From the perspective of cost, researchers have focused on various issues. In \cite{eismann2020predicting}, the authors present a technique that predicts the cost of function workflows. The authors in \cite{mahajanoptimal2019} propose an algorithm that optimizes the cost of function workflow through function fusion and placement. In \cite{elgamal2018costless} the authors have identified different operation regimes that optimizes the cost of both customer and provider. From the perspective of resource management, researchers have mainly focused on runtime CPU allocation considering the QoS \cite{hoseinyfarahabady2017qos,kim2018dynamic}.

The authors in \cite{wang2018peeking} and \cite{lloyd2018serverless} are among the firsts to investigate function latency considering cold and warm states. In recent works, researchers are also proposing different solutions to this problem. In \cite{lloyd2018improving}, the authors have addressed cold starts from the end user perspective and mitigated it by periodically sending low cost service requests. The authors in \cite{mohan2019agile} have pre-initialized resources, like networking elements, and associated them with containers as required. In \cite{bermbach2020using}, the authors have provisioned containers in advance by leveraging function composition knowledge. The authors in \cite{shahrad2020serverless} propose a window based approach to load or unload functions by analyzing their invocation patterns. However, none of these works modeled function memory consumption and response time, making them inapplicable in capacity planning.

\section{Conclusion and Future Work}
\label{sec:conc}
We have presented COCOA, a cold-start aware sizing method for on-premise FaaS platforms. 
COCOA leverages an LQN model and M/M/k setup models to obtain different performance estimates and consequently, predict the required system capacity. We have illustrated the improvements yielded by COCOA with multiple experiments, showing that COCOA can help in provisioning FaaS systems that satisfy SLAs .

A future research direction could be incorporating burstiness in the workload that triggers more resource intensive actions and dealing with autoscaling scenario where multiple function replicas need to be instantiated.




\begin{thebibliography}{10}
	\providecommand{\url}[1]{#1}
	\csname url@samestyle\endcsname
	\providecommand{\newblock}{\relax}
	\providecommand{\bibinfo}[2]{#2}
	\providecommand{\BIBentrySTDinterwordspacing}{\spaceskip=0pt\relax}
	\providecommand{\BIBentryALTinterwordstretchfactor}{4}
	\providecommand{\BIBentryALTinterwordspacing}{\spaceskip=\fontdimen2\font plus
		\BIBentryALTinterwordstretchfactor\fontdimen3\font minus
		\fontdimen4\font\relax}
	\providecommand{\BIBforeignlanguage}[2]{{%
			\expandafter\ifx\csname l@#1\endcsname\relax
			\typeout{** WARNING: IEEEtran.bst: No hyphenation pattern has been}%
			\typeout{** loaded for the language `#1'. Using the pattern for}%
			\typeout{** the default language instead.}%
			\else
			\language=\csname l@#1\endcsname
			\fi
			#2}}
	\providecommand{\BIBdecl}{\relax}
	\BIBdecl
	
	\bibitem{wang2018peeking}
	L.~Wang \emph{et~al.}, ``{Peeking Behind the Curtains of Serverless
		Platforms},'' in \emph{Proc. of USENIX ATC}, 2018, pp. 133--146.
	
	\bibitem{jabbari2018towards}
	R.~Jabbari \emph{et~al.}, ``{Towards a benefits dependency network for DevOps
		based on a systematic literature review},'' \emph{Journal of Software:
		Evolution and Process}, vol.~30, no.~11, p. e1957, 2018.
	
	\bibitem{lloyd2018serverless}
	W.~Lloyd \emph{et~al.}, ``{Serverless Computing: An Investigation of Factors
		Influencing Microservice Performance},'' in \emph{Proc. of IC2E}.\hskip 1em
	plus 0.5em minus 0.4em\relax IEEE, 2018, pp. 159--169.
	
	\bibitem{basu2018adaptive}
	S.~Basu \emph{et~al.}, ``{Adaptive TTL-Based Caching for Content Delivery},''
	\emph{{Trans. on Networking}}, vol.~26, no.~3, pp. 1063--1077, 2018.
	
	\bibitem{che2002hierarchical}
	H.~Che, Y.~Tung, and Z.~Wang, ``{Hierarchical Web Caching Systems: Modeling,
		Design and Experimental Results},'' \emph{IEEE JSAC}, vol.~20, no.~7, pp.
	1305--1314, 2002.
	
	\bibitem{franks2008enhanced}
	G.~Franks \emph{et~al.}, ``{Enhanced Modeling and Solution of Layered Queueing
		Networks},'' \emph{Trans. on Soft. Eng.}, vol.~35, no.~2, pp. 148--161, 2008.
	
	\bibitem{gandhi2013exact}
	A.~Gandhi \emph{et~al.}, ``{Exact Analysis of the M/M/k/setup Class of Markov
		Chains via Recursive Renewal Reward},'' in \emph{Proc. of SIGMETRICS}.\hskip
	1em plus 0.5em minus 0.4em\relax ACM, 2013, pp. 153--166.
	
	\bibitem{fricker2012versatile}
	C.~Fricker, P.~Robert, and J.~Roberts, ``{A Versatile and Accurate
		Approximation for LRU Cache Performance},'' in \emph{Proc. of ITC}.\hskip 1em
	plus 0.5em minus 0.4em\relax IEEE, 2012, pp. 1--8.
	
	\bibitem{dehghan2019utility}
	M.~Dehghan \emph{et~al.}, ``{A Utility Optimization Approach to Network Cache
		Design},'' \emph{Trans. on Networking}, vol.~27, no.~3, pp. 1013--1027, 2019.
	
	\bibitem{bertoli2009jmt}
	M.~Bertoli, G.~Casale, and G.~Serazzi, ``{JMT: Performance Engineering Tools
		for System Modeling},'' \emph{{ACM SIGMETRICS PER}}, vol.~36, no.~4, pp.
	10--15, 2009.
	
	\bibitem{glassman1994caching}
	S.~Glassman, ``A caching relay for the world wide web,'' \emph{Computer
		Networks and ISDN Systems}, vol.~27, no.~2, pp. 165--173, 1994.
	
	\bibitem{mikhail2019comparison}
	M.~Shilkov, ``{Comparison of Cold Starts in Serverless Functions across AWS,
		Azure, and GCP},'' \url{https://mikhail.io/serverless/coldstarts/big3}, 2019,
	{Accessed: 2020-06-30}.
	
	\bibitem{akkus2018sand}
	I.~Akkus \emph{et~al.}, ``{SAND: Towards High-Performance Serverless
		Computing},'' in \emph{Proc. of USENIX ATC}, 2018, pp. 923--935.
	
	\bibitem{LatoucheR99}
	G.~Latouche and V.~Ramaswami, \emph{Introduction to Matrix Analytic Methods in
		Stochastic Modeling}.\hskip 1em plus 0.5em minus 0.4em\relax SIAM, 1999.
	
	\bibitem{riska2002mamsolver}
	A.~Riska and E.~Smirni, ``{MAMSolver: A Matrix Analytic Methods Tool},'' in
	\emph{Proc. of Modelling Techniques and Tools for Computer Performance
		Evaluation}.\hskip 1em plus 0.5em minus 0.4em\relax Springer, 2002, pp.
	205--211.
	
	\bibitem{riska2007etaqa}
	A.~Riska and E.~Smirni, ``{ETAQA Solutions for Infinite Markov Processes with
		Repetitive Structure},'' \emph{INFORMS Journal on Computing}, vol.~19, no.~2,
	pp. 215--228, 2007.
	
	\bibitem{tribastone2010performance}
	M.~Tribastone, P.~Mayer, and M.~Wirsing, ``Performance prediction of
	service-oriented systems with layered queueing networks,'' in \emph{Proc. of
		ISoLA}.\hskip 1em plus 0.5em minus 0.4em\relax Springer, 2010, pp. 51--65.
	
	\bibitem{gias2019atom}
	A.~U. Gias, G.~Casale, and M.~Woodside, ``{ATOM: Model-Driven Autoscaling for
		Microservices},'' in \emph{Proc. of ICDCS}.\hskip 1em plus 0.5em minus
	0.4em\relax IEEE, 2019, pp. 1994--2004.
	
	\bibitem{harchol2013performance}
	M.~Harchol-Balter, \emph{{Performance Modeling and Design of Computer Systems:
			Queueing Theory in Action}}.\hskip 1em plus 0.5em minus 0.4em\relax Cambridge
	Univ. Press, 2013.
	
	\bibitem{shousha1998applying}
	C.~Shousha \emph{et~al.}, ``Applying performance modelling to a
	telecommunication system,'' in \emph{Proc. of WOSP}.\hskip 1em plus 0.5em
	minus 0.4em\relax ACM, 1998, pp. 1--6.
	
	\bibitem{casale2019automated}
	G.~Casale, ``{Automated Multi-paradigm Analysis of Extended and Layered
		Queueing Models with LINE},'' in \emph{Comp. Proc. of ICPE}.\hskip 1em plus
	0.5em minus 0.4em\relax ACM/SPEC, 2019, pp. 37--38.
	
	\bibitem{casale2018analyzing}
	G.~Casale, ``Analyzing replacement policies in list-based caches with
	non-uniform access costs,'' in \emph{Proc. of INFOCOM}.\hskip 1em plus 0.5em
	minus 0.4em\relax IEEE, 2018, pp. 432--440.
	
	\bibitem{nelson2013probability}
	R.~Nelson, \emph{{Probability, Stochastic Processes, and Queueing Theory: The
			Mathematics of Computer Performance Modeling}}.\hskip 1em plus 0.5em minus
	0.4em\relax Springer, 2013.
	
	\bibitem{martina2014unified}
	V.~Martina, M.~Garetto, and E.~Leonardi, ``A unified approach to the
	performance analysis of caching systems,'' in \emph{Proc. of INFOCOM}.\hskip
	1em plus 0.5em minus 0.4em\relax IEEE, 2014, pp. 2040--2048.
	
	\bibitem{AWSLambdaLimit}
	``{AWS Lambda limits},''
	\url{https://docs.aws.amazon.com/lambda/latest/dg/gettingstarted-limits.html},
	{Accessed: 2020-06-30}.
	
	\bibitem{eismann2020predicting}
	S.~Eismann \emph{et~al.}, ``{Predicting the Costs of Serverless Workflows},''
	in \emph{Proc. of ICPE}.\hskip 1em plus 0.5em minus 0.4em\relax ACM/SPEC,
	2020, pp. 265--276.
	
	\bibitem{mahajanoptimal2019}
	K.~Mahajan \emph{et~al.}, ``{Optimal Pricing for Serverless Computing},'' in
	\emph{Proc. of GLOBECOM}.\hskip 1em plus 0.5em minus 0.4em\relax IEEE, 2019.
	
	\bibitem{elgamal2018costless}
	T.~Elgamal \emph{et~al.}, ``{Costless: Optimizing Cost of Serverless Computing
		through Function Fusion and Placement},'' in \emph{Proc. of SEC}.\hskip 1em
	plus 0.5em minus 0.4em\relax IEEE/ACM, 2018, pp. 300--312.
	
	\bibitem{hoseinyfarahabady2017qos}
	M.~R. HoseinyFarahabady \emph{et~al.}, ``{A QoS-Aware Resource Allocation
		Controllerfor Function as a Service (FaaS) Platform},'' in \emph{Proc. of
		ICSoC}.\hskip 1em plus 0.5em minus 0.4em\relax Springer, 2017, pp. 241--255.
	
	\bibitem{kim2018dynamic}
	Y.~K. Kim \emph{et~al.}, ``{Dynamic Control of CPU Usage in a Lambda
		Platform},'' in \emph{Proc. of CLUSTER}.\hskip 1em plus 0.5em minus
	0.4em\relax IEEE, 2018, pp. 234--244.
	
	\bibitem{lloyd2018improving}
	W.~Lloyd \emph{et~al.}, ``{Improving Application Migration to Serverless
		Computing Platforms: Latency Mitigation with Keep-Alive Workloads},'' in
	\emph{Com. Proc. of UCC}.\hskip 1em plus 0.5em minus 0.4em\relax IEEE/ACM,
	2018, pp. 195--200.
	
	\bibitem{mohan2019agile}
	A.~Mohan \emph{et~al.}, ``{Agile Cold Starts for Scalable Serverless},'' in
	\emph{Proc. of HotCloud}.\hskip 1em plus 0.5em minus 0.4em\relax USENIX,
	2019, pp. 1--6.
	
	\bibitem{bermbach2020using}
	D.~Bermbach, A.-S. Karakaya, and S.~Buchholz, ``{Using Application Knowledge to
		Reduce Cold Starts in FaaS Services},'' in \emph{Proc. of SAC}.\hskip 1em
	plus 0.5em minus 0.4em\relax ACM, 2020, pp. 134--143.
	
	\bibitem{shahrad2020serverless}
	M.~Shahrad \emph{et~al.}, ``{Serverless in the Wild: Characterizing and
		Optimizing the Serverless Workload at a Large Cloud Provider},'' in
	\emph{Proc. of USENIX ATC}, 2020.
	
\end{thebibliography}
\end{document}